\documentstyle[epsf]{mn}  
\newcommand{\beq}{\begin{eqnarray}}
\newcommand{\eeq}{\end{eqnarray}}

\newcommand{\bfq}{{\bf q}}
\newcommand{\bfr}{{\bf r}}
\newcommand{\bfv}{{\bf v}}
\newcommand{\bfk}{{\bf k}}
\newcommand{\bfL}{{\bf L}}
\newcommand{\ti}{{\times}}

\begin{document}   
\voffset -0.3truecm
\title[Effects of merging histories on angular momentum
distribution]{
\vglue -1.5truecm{
%\rightline{\Large To appear in {\em Mon. Not. R. Astron. Soc. }}
%\vglue -0.25truecm
{\rightline{\Large\bf OU-TAP 62}}
\vglue 1.5truecm}
\noindent 
Effects of merging histories on angular momentum
distribution of dark matter haloes}

%\title[Effects of merging histories on angular momentum distribution]
%{Effects of merging histories on angular momentum distribution of dark
%matter haloes}

\author[M. Nagashima \& N. Gouda]{Masahiro   
Nagashima\thanks{Research Fellow of the Japan Society for the   
Promotion of Science} and Naoteru Gouda\\    
Department of Earth and Space Science, Graduate School of Science,
Osaka University, Toyonaka, Osaka 560-0043, Japan\\    
Email: masa, gouda@vega.ess.sci.osaka-u.ac.jp}   
%%%%%%%%%%%%%%%%%%%%%%%%%%%%%%%%%%%%%%%%%%%%%%%%%%%%%%%%%%%%%   
\maketitle   
   
\begin{abstract}   
The effects of merging histories of proto-objects on the angular
momentum distributions of the present-time dark matter haloes are
analysed.  An analytical approach to the analysis of the angular
momentum distributions assumes that the haloes are initially
homogeneous ellipsoids and that the growth of the angular momentum of
the haloes halts at their maximum expansion time.  However, the
maximum expansion time cannot be determined uniquely, because in the
hierarchical clustering scenario each progenitor, or subunit, of the
halo has its own maximum expansion time.  Therefore the merging
history of the halo may be important in estimating its angular
momentum.  Using the merger tree model by Rodrigues \& Thomas, which
takes into account the spatial correlations of the density
fluctuations, we have investigated the effects of the merging
histories on the angular momentum distributions of dark matter haloes.
It was found that the merger effects, that is, the effects of the
inhomogeneity of the maximum expansion times of the progenitors which
finally merge together into a halo, do not strongly affect the final
angular momentum distributions, so that the homogeneous ellipsoid
approximation happens to be good for the estimation of the angular
momentum distribution of dark matter haloes.  This is because the
effect of the different directions of the angular momenta of the
progenitors cancels out the effect of the inhomogeneity of the maximum
expansion times of the progenitors.

The contribution of the orbital angular momentum to the total angular
momentum when two or more pre-existing haloes merge together, was also
investigated.  It is shown that this contribution is more important
than that of the angular momentum of diffuse accreting matter to the
total angular momentum, especially when the mergers occur many times.
\end{abstract}   
   

\section{INTRODUCTION}   
Galaxy formation is one of the most important problems in
astrophysics.  In general, it includes different physical processes
such as the evolution of cosmological fluctuations, star formation,
heating processes of baryonic gas from supernovae, dynamical and
chemical evolution of gas, etc. Hence it is difficult to attack the
problem of the formation and evolution of galaxies in a way that
connects all of the complicated physical processes.

One of the important keys to understanding galaxy formation is the
angular momentum of dark matter haloes.  For example, there is a
strong correlation between the angular momentum and the morphology of
galaxies.  The mean value of the angular momentum of ellipticals is
much smaller than that of spirals.  Thus we may expect to understand
the origin of the morphological distinction from analysis of the
angular momentum.  In the standard model, it is considered that dark
matter dominates in our Universe and that galaxies and clusters of
galaxies have been formed by the gravitational growth of small initial
density fluctuations.  The dark matter has collapsed and virialized by
self-gravitational instability into objects which are called `dark
matter haloes' or `dark haloes'.  The large haloes are generally
considered to have been formed hierarchically by the clustering of
smaller ones (so-called `hierarchical clustering').  During this
process, each dark halo obtains its angular momentum by tidal forces
(Hoyle 1949; Peebles 1969; White 1984; Barnes \& Efstathiou 1987).
Also, some properties of the baryonic component of galaxies e.g.
morphologies, surface brightness and so on, have been discussed as a
function of angular momentum of dark haloes (e.g., Dalcanton, Spergel
\& Summers 1997; Jimenez et al. 1997; Mo, Mao \& White 1998).

Doroshkevich (1970) and White (1984) found that the angular momentum
evolves proportionally to $t$ ($t$ is the cosmic time) during the
linear regime, considering the initially homogeneous density ellipsoid
as a proto-object in the Einstein-de Sitter universe.  This prediction
has been confirmed by the N-body simulations by White (1984) and
Barnes \& Efstathiou (1987).  It should be noted that White (1984) did
not consider whether the dark matter in the region under consideration
would in fact grow into a collapsed object or not in his analytical
prediction.

In the analysis of the evolution of the angular momentum of a
proto-object, it is assumed that the dark matter around a density peak
maximum collapses into a halo (Hoffman 1986, 1988; Ryden \& Gunn 1987;
Ryden 1988; Caimmi 1989; Quinn \& Binney 1992 and Eisenstein \& Loeb
1995a, b).  The angular momentum distributions were derived
analytically by Heavens \& Peacock (1988), 
%Steinmetz \& Bartelmann(1992), 
and Catelan \& Theuns (1996a, b).  They used the peak formalism of
Gaussian random fields (Peacock \& Heavens 1985; Bardeen et al. 1986),
in order to analyse statistically a density profile around the density
peak maximum.  Catelan \& Theuns also considered the case of
non-Gaussian initial conditions (Catelan \& Theuns 1997).  Susa,
Sasaki \& Tanaka investigated the angular momentum distribution by
using the Press-Schechter formalism (Press \& Schechter 1974) instead
of the peak formalism.

It should be noted that when the distribution of the angular momentum
was derived, there was assumed that an `object' is the initially
homogeneous density ellipsoid and that the acquisition of the angular
momentum halts at the maximum expansion time of the object (Heavens \&
Peacock 1988; Catelan \& Theuns 1996a, b).  The maximum expansion time
is usually estimated from the averaged density contrast of the object
by assuming spherically symmetric collapse.  In spherical collapse,
the averaged linear density contrast of the object reaches to $1.05$
at the maximum expansion time (e.g. Peebles 1993).  However, in the
hierarchical clustering scenario, small objects collapse firstly.
Some of the small objects may then merge together, leading to the
present-time large halo.  It is general in this scenario that the
acquisition of the angular momentum of a dominant proto-object of a
halo may halts earlier than the maximum expansion time estimated
through the averaged density contrast of the present-time halo.
Hence, if we believe in the merging history of the proto-objects, then
we should take into account the halting time of each proto-object, a
contribution of the orbital angular momentum of the merging
proto-object, and the angular momentum of the accreting matter.

Merging history models of dark haloes have been used for the
semi-analytical methods of galaxy formations (Cole \& Kaiser 1988;
Kauffmann \& White 1993; Rodrigues \& Thomas 1996; Roukema et
al. 1997).  Cole \& Kaiser developed the block model, considering the
one-point distribution function of density fluctuations.  Therefore
spatial correlations of the density fields were not taken into
account.  Kauffmann \& White (1993) extended the Press-Schechter
formalism and constructed the merging histories by coupling this
formalism with the Monte Carlo method.  The extension in question is
mainly based on the results of Bower (1991).  Kauffmann \& White
(1993) have utilized the one-point distribution function as well.  On
the other hand, Yano, Nagashima \& Gouda (1996) have shown that the
spatial correlation of the density field is important when calculating
mass function of dark haloes.  The analytically derived mass function
accounting explicitly for the two-point correlation function, differs
significantly from that found by the Press-Schechter mass function.
So we believe that the merging histories should also include spatial
information.  Rodrigues \& Thomas (1996) constructed the merging
history model (which we call `the merging cell model', for
convenience), including the information on the spatial correlations.
In the merging cell model, the random Gaussian density fluctuation
field is realized on spatial grids as in the construction of the
initial conditions of N-body simulations.  Therefore this model
includes, naturally, information on the spatial correlation.  Then, by
finding the collapsed region whose linear density contrast, $\delta$,
averaged over the region is $\delta_{c}=1.69$ ($\delta_{c}$ is the
critical density contrast for collapse, see, e.g., Peebles 1993), we
can construct merger trees.  We expect that this model is more
realistic and useful in describing the galaxy formations although a
spherical collapse of each block is assumed.  Nagashima \& Gouda
(1997) have analysed this merging cell model by comparing the mass
functions given by the merging cell model with the mass functions by
Yano, Nagashima \& Gouda (1996).  They found that both types of the
mass functions are in good agreement.

In this paper, using the merging cell model, we calculate both the
density fields and the velocity fields in order to estimate the
angular momentum distribution of dark haloes.  In a similar way to the
previous analytical work (Heavens \& Peacock 1988; Catelan \& Theuns
1996a, b), the initial angular momentum of each halo is calculated
through the velocity field, then the angular momentum of each halo is
evolved proportionally to the cosmic time $t$.  Since the overdensity
of each halo is known, we can estimate its maximum expansion time.  By
using this time, we compute the angular momentum of the halo.  When a
merger between the haloes occurs, we investigate the spins and the
orbital angular momentum of these haloes around the centre-of-mass of
a larger new common halo, and the angular momentum of the matter being
accreted into the new halo.  The sum of these angular momentum
components is the spin of the new halo.  This process is repeated when
the mergers occur.  It should be noted that the block model and the
extension of the Press-Schechter formalism cannot deal with angular
momentum, because these models ignore the velocity field structures.

To compare our method with the previous work, we also calculate the
angular momentum of homogeneous density haloes by smoothing density
contrasts in each region occupied by haloes at the present epoch.  In
this way we investigate the effects of the merging histories by
showing the angular momentum distribution of dark haloes.

Our aim is to clarify the physical effects of the merging histories of
dark haloes, namely, the effects of the difference in the maximum
expansion times and the difference in the directions of the angular
momenta of the proto-objects, that is, the subunits of the haloes.  An
advantage of our method is that it makes available a qualitative
analysis of the acquisition of angular momentum.  Of course, we must
perform N-body simulations in order to understand the process
quantitatively (Efstathiou \& Jones 1979; Barnes \& Efstathiou 1987;
Warren et al. 1992; Lemson \& Kauffmann 1997), but our calculations
require less computing time than do N-body simulations.  Therefore, we
can investigate many model parameters for a short time.  We believe
that the semi-analytical methods are complementary to N-body
simulations.

In Section 2, previous analytical work is reviewed briefly.  In
Section 3, the merging cell model is characterized.  In Section 4, the
method of calculating the angular momenta of haloes in the merging
cell model is shown.  In Section 5, we consider the angular momentum
distribution.  We show the merger effects as well as the role of the
orbital angular momentum of the merging objects in their common halo.
Section 6 is devoted to conclusions and discussion.

%%%%%%%%%%%%%%%%%%%%%%%%%%%%%%%%%%% section 2 %%%%%%%%%%%%%%%%%%%%%%%%%   
\section{TIME DEPENDENCE OF ANGULAR MOMENTUM} 

In this section, we review the evolution of angular momentum according
to the White's method (White 1984).  The notations are almost the same
as in Catelan \& Theuns (1996a).

The angular momentum $\bfL$ of an object occupying the volume $V$ in
the Eulerian space is defined as
\begin{eqnarray}
\bfL(t)\equiv\int_{V}\rho(\bfr)(\bfr-\bfr_{0})\ti(\bfv-\bfv_{0}) d\bfr,
\end{eqnarray}
where $\rho$ is the density, $\bfr$ is the Eulerian coordinate, $\bfv$
is the velocity, and the subscript $0$ denotes the centre-of-mass of
the volume $V$.  This transforms to the Lagrangian description,
\begin{eqnarray}
\bfL&=&\int_{\Gamma}\eta_{0}a^{2}\left[\bfq-\bfq_{0}
+D\nabla(\phi-\phi_{0})\right]
\ti\dot{D}\nabla(\phi-\phi_{0}) d\bfq\nonumber\\
&=&\int_{\Gamma}\eta_{0}a^{2}\dot{D}(\bfq-\bfq_{0})
\ti\nabla(\phi-\phi_{0}) d\bfq,\label{eqn:ell}
\end{eqnarray}
with $\eta_{0}=\bar\rho a^{3}$, where $\eta_{0}$ is the comoving mean
density of the universe, $\bar\rho$ is the mean density of the
universe, $a$ is the cosmic scale factor, $\bfq$ is the Lagrangian
coordinate, $\phi$ is proportional to the gravitational potential in
the Einstein-de Sitter universe, $D$ is the linear growth factor
($D=a$ in the Einstein-de Sitter universe), and $\Gamma$ denotes the
Lagrangian volume corresponding to $V$.  The dot represents time
derivative, $d/dt$.  In the above transformation, we use the
Zel'dovich approximation (Zel'dovich 1970),
\begin{eqnarray} 
\bfr(t,\bfq)=a(t)[\bfq+D(t)\nabla\phi(\bfq)]. 
\end{eqnarray}
Thus we find the growth rate of the angular momentum as 
\begin{eqnarray} 
L\propto a^{2}\dot{D}\propto t \label{eqn:lt}
\end{eqnarray} 
in the Einstein-de Sitter universe.

The growth of the angular momentum halts at the maximum expansion time
of the object.  In the previous work, this maximum expansion time is
estimated via the density contrast at a centre of the object with the
Lagrangian volume $\Gamma$, which is smoothed with a filtering scale
corresponding to the mass of the object.  Using the spherical collapse
approximation, the maximum expansion time $t_{m}$ is obtained by
$t_{m}=t_{0}(1.05/\delta)^{3/2}$, where $t_{0}$ is the present time
and $\delta$ is the density contrast at the centre of $\Gamma$.

Note that the above discussion is justified only when the Zel'dovich
approximation is valid anywhere in $\Gamma$ until $t_{m}$.  However,
in hierarchical clustering scenarios, collapsed haloes cluster and
grow up gradually, so that the Zel'dovich approximation in such
pre-existing haloes is broken earlier than $t_{m}$ of the final
collapsed halo with $\Gamma$.  Therefore, we need to investigate the
case in which the growth of angular momenta of the pre-existing haloes
halts earlier than the maximum expansion time of the final collapsed
halo which is formed via mergers of the pre-existing haloes.  In this
paper, a {\it progenitor} means such a pre-existing halo.

%%%%%%%%%%%%%%%%%%%%%%%%%%%%%%% section 3 %%%%%%%%%%%%%%%%%%%
\section{MERGING CELL MODEL}   
Let us describe the general aspects of the merging cell model in
accord with Nagashima \& Gouda (1997).

First, the random Gaussian density field is realised in a periodic
cubical box of side $L$.  The Fourier mode of the density contrast
$\delta [=(\rho-\bar\rho)/\bar\rho]$ obeys the following probability
distribution function for its amplitude and phase. In the random
Gaussian distribution, (Bardeen et al. 1986),
\begin{equation}   
P(\vert\delta_{\bf k}\vert,\phi_{\bf k})d\vert\delta_{\bf k}\vert
d\phi_{\bf k}=\frac{2\vert\delta_{\bf k}\vert}{P(k)}\exp\left\{-\frac
{ \vert\delta_{\bf k}\vert^{2}}{P(k)}\right\}d\vert\delta_{\bf
k}\vert\frac{d\phi_{\bf k}}{2\pi}\label{eqn:power},
\end{equation}   
where $\phi_{\bf k}$ is the random phase of $\delta_{\bf k}$,
$\delta_{\bf k}=|\delta_{\bf k}|\exp{(i\phi_{\bf k})}$, and $P(k)$ is
the power spectrum.  Then, the density contrast at each grid (`cell')
is given by the Fourier transform,
\begin{equation}   
\delta({\bf q})=\frac{V}{(2\pi)^{3}}\int_{0}^{k_{c}}\delta_{\bf   
k}e^{i\bf k\cdot q}d^{3}k,   
\end{equation}   
where $k_{c}$ is the cut-off wavenumber corresponding to the cell
size.
   
Next, the fluctuations of the various smoothing levels are constructed
by averaging the density fluctuations within cubical {\it blocks} of
side 2, 4, \ldots, $L_{\mbox{box}}$.  At each smoothing level,
displacing the smoothing grids by half a blocklength along the
coordinate axes, eight sets of {\it overlapping grids} are constructed
to centre approximately the density peak within one of them.
   
Then, the density fluctuations within blocks and cells are combined
into a single list and ordered by decreasing density.  The
fluctuations are investigated from the top of this list.  The
following criteria decide whether a block or a cell can collapse.
Note the terminology that a {\it halo} is a block or a cell that has
already collapsed, and an {\it investigating} region is a block or a
cell whose linear density contrast is just equal to $\delta_c$ at the
reference time.
\begin{enumerate}   
\renewcommand{\labelenumi}{(\alph{enumi})}
\item If the investigating region includes no haloes, its can collapse
and can be identified as a new halo.
\item When the investigating region includes a part of a halo, if the
{\it overlapping region} has at least half of the minimum of the
masses of the halo and the investigating region, then the
investigating region can collapse.  This is the criterion of collapse
of the investigating region and merging for the overlapped haloes. We
call this criterion as {\it the overlapping criterion}.
\item If the investigating region has half or more of its mass
contained in exactly one pre-existing halo and the overlapping
criterion (b) is satisfied for any overlapped haloes then they can be
merged together as part of a new structure.  This is the criterion
for linking of haloes.
\end{enumerate}   
   
These criteria were chosen by Rodrigues \& Thomas (1996).  We
introduce the overlapping parameters $x$ and $y$, according to
Nagashima \& Gouda (1997).  The parameter $x$ means the ratio of the
mass of the overlapping region to the lesser of the masses of the halo
and the investigating region.  Then, the Rodrigues \& Thomas's
criterion corresponds to $x=1/2$.  The parameter $y$ quantifies the
linking criterion.  Using this parameter $y$, we change the criterion
(c) as follows: If the investigating region has $y$ times ($x\leq
y\leq 1/2$) or more of its mass contained in exactly one pre-existing
halo and the overlapping criterion (b) is satisfied for any overlapped
haloes then they can be merged together as part of a new structure.

In the case of $x=y=1/2$, it is shown by Nagashima and Gouda (1997)
that the mass functions given by the merging cell model are in
agreement with those of Yano, Nagashima and Gouda (1996), whose
formula implicitly includes an overlapping criterion which corresponds
to $x=y=1/2$ as an assumption.  Thus this agreement is lost in the
case of $x=1/8$ and $y=1/2$.  However, we need more detailed
investigations by, e.g., N-body simulations, in order to decide
realistic values of $x$ and $y$.  Therefore we use two parameter sets
$x=y=1/2$, and $x=1/8$ and $y=1/2$ in this paper.

%%%%%%%%%%%%%%%%%%%%%%%%%%%% METHOD %%%%%%%%%%%%%%%%%%%%%%%%%
\section{METHOD OF CALCULATION OF ANGULAR MOMENTUM}   
In this section we describe the method of calculation of the angular
momentum of haloes.

From the Fourier component $\delta_{\bfk}$ obtained by the Monte Carlo
method using eq.(\ref{eqn:power}) and the Poisson equation
\begin{eqnarray}
\nabla^{2}\phi=-\frac{\delta}{D_{0}},
\end{eqnarray}
we can estimate the velocity $\bfv$ and potential $\phi$ at each cell,
\begin{eqnarray}
\phi({\bfq})&=&\frac{V}{(2\pi)^{3}D_{0}}\int\frac{\delta_{\bfk}}{k^{2}}
e^{i\bf k\cdot q}d^{3}k,\\
\bfv({\bfq})&=&i\frac{a_{0}\dot{D_{0}}V}{(2\pi)^{3}D_{0}}\int\frac{{\bfk}\delta_{\bf
k}}{k^{2}}e^{i\bf k\cdot q}d^{3}k,
\end{eqnarray}
where $a_{0}$ and $D_{0}$ are the scale factor and the growing mode at
the present epoch.

When a halo collapses, we divide this new halo's angular momentum into
three components, which are, (1) the sum of the spin angular momenta
of pre-existing haloes $\bfL_{spin}$, (2) the sum of the orbital
angular momenta of pre-existing haloes $\bfL_{orb}$, and (3) the
angular momentum of the matter accreting into a new halo,
$\bfL_{acc}$.  In order to calculate the above quantities, we need the
mass of the pre-existing haloes, and the positions and velocities of
the center-of-mass of both the pre-existing haloes and the new common
halo.  The growth of the angular momenta of the orbital and accreting
matter is proportional to $t$, according to eq.(\ref{eqn:lt}).  These
components grow until the maximum expansion time of the new common
halo.  Therefore, we need the density contrast of the new common halo
in order to estimate the maximum expansion time.  We estimate the
relation between the linear density contrast at the present epoch and
the maximum expansion time $t_{m}$ using the spherical collapse model,
\begin{eqnarray}
t_{m}=t_{0}\left(\frac{1.05}{\delta}\right)^{\frac{3}{2}}.
\label{eqn:max}
\end{eqnarray}

\begin{figure}
\epsfxsize=8cm
\epsfbox{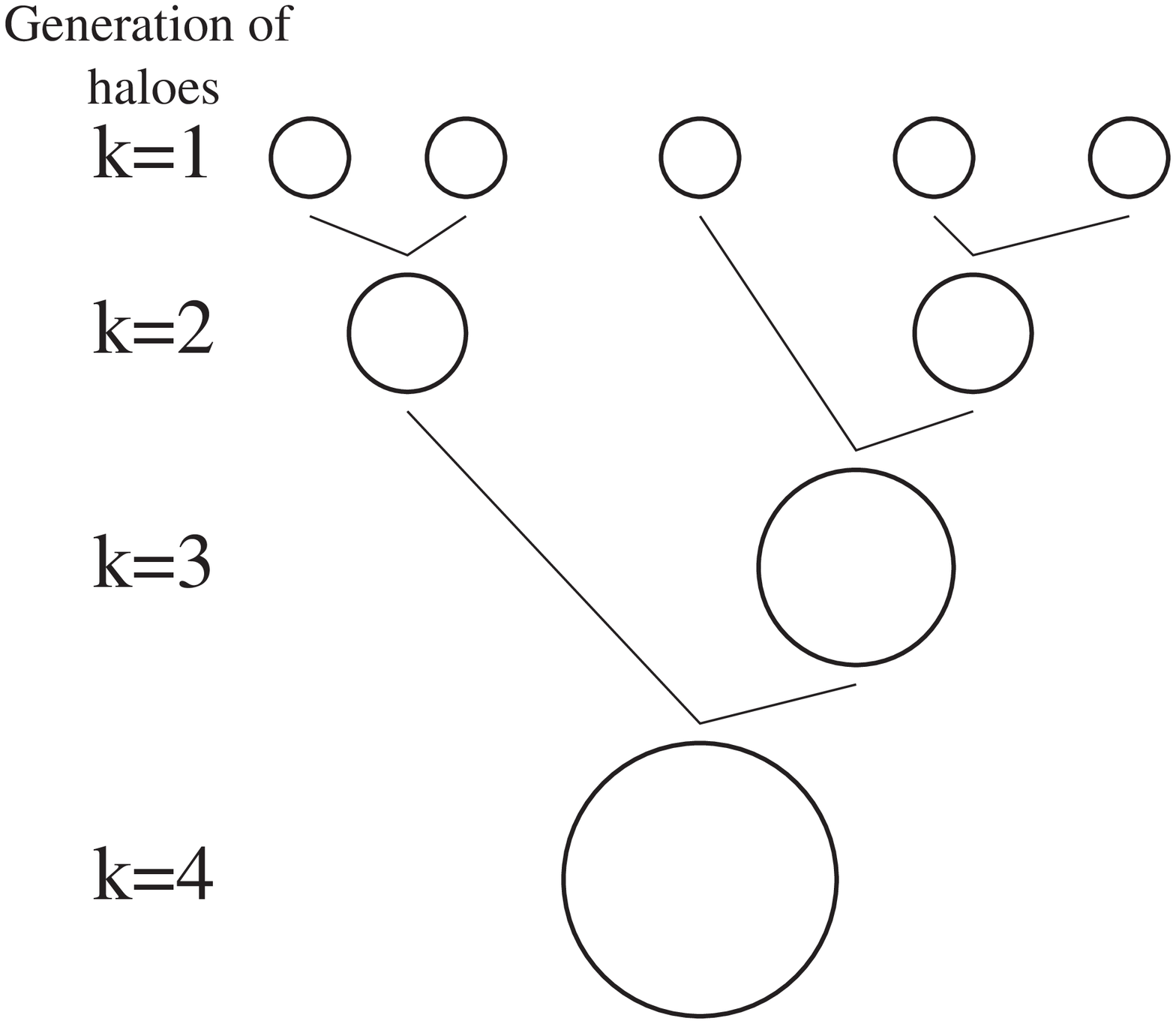}
\caption{Example of generation and merging history of haloes.  The
first generation haloes include no progenitors.  The final halo
belongs to the fourth generation, and is formed by the merger of the
third generation halo and the second generation halo.  We omit the
accreting matter in this figure.}
\label{fig:merge}
\end{figure}

Now we define the generation of haloes as follows (see
Fig.\ref{fig:merge}).  A collapsed block which has no progenitors is
defined as a first generation halo $(k=1)$.  The $k$-th generation
halo is defined as a halo which includes the $(k-1)$-th generation
pre-existing haloes whose number specifying their generations is the
largest of its progenitors.  If the new halo is a $k$-th generation
halo ($k\geq 2$), these three components of angular momentum are
expressed as
\begin{eqnarray}
\bfL_{spin}&\equiv&\sum_{i}\bfL_{spin,i}^{k'(i)}~~~[1\leq k'(i)\leq
k-1],\\ 
\bfL_{orb}&\equiv&\sum_{i}M_{i}(\bfq_{i}-\bfq_{CM})\times
(\bfv_{i}-\bfv_{CM})\left(\frac{1.05}{\delta^{k}}\right)^{\frac{3}{2}},
\\ \bfL_{acc}&\equiv&M_{cell}\sum_{j}(\bfq_{j}-\bfq_{CM})\times
(\bfv_{j}-\bfv_{CM})
\left(\frac{1.05}{\delta^{k}}\right)^{\frac{3}{2}}\label{eqn:acc},
\end{eqnarray}
where the subscripts $i$ and $j$ stand for the isolated pre-existing
haloes just before the new common halo collapses and cells of
accreting matter respectively, so that $\bfL_{spin,i}^{k'(i)}, M_{i},
\bfq_{i}$ and $\bfv_{i}$ are respectively the spin, mass, position and
velocity of the centre-of-mass of the $i$-th progenitor.  The
superscripts $k$ and $k'(i)$ show the generation of haloes.  The
isolated $k'(i)$-th generation pre-existing haloes $[k'(i)\geq 1]$
merge together into the $k$-th generation halo.  The subscript $CM$
denotes the centre-of-mass of the new halo, $M_{cell}$ is a mass of
one cell, and $\delta^{k}$ is associated with a new collapsed $k$-th
generation halo, because the angular momenta of the orbital and
accreting matter grow until the maximum expansion time of the new
collapsed halo.  Finally, we obtain a total angular momentum
$\bfL_{spin}^{k}$ of the new collapsed halo by summing the above
quantities,
\begin{eqnarray}
\bfL_{spin}^{k}=\bfL_{spin}+\bfL_{orb}+\bfL_{acc}.\label{eqn:lnew}
\end{eqnarray}

If this new collapsed $k$-th generation halo will merge again into a
collapsed halo at a later epoch, $\bfL_{spin}^{k}$ is used as
$\bfL_{spin,i}^{k}$ in the estimation of the angular momentum
$\bfL^{k+1}$ of the later collapsed $(k+1)$-th generation halo.  Note
that when a first ($k=1$) generation halo collapses, this halo has no
progenitors, so this halo is formed by only the accreting matter.
Therefore the angular momentum of this halo is estimated by only
eq.(\ref{eqn:acc}), i.e., $\bfL_{spin}^{1}=\bfL_{acc}$.  This
estimation is the same as in the homogeneous collapse,
eq.(\ref{eqn:ell}).

When considering a homogeneous case as in the previous analytical work
(Heavens \& Peacock 1988; Catelan \& Theuns 1996a, b), we first
calculate the regions of collapsed haloes as described in Section 3.
Then we average the density contrasts of cells in each halo, and
obtain the averaged density contrast $\delta_{avg}$.  Finally, using
only eq.(\ref{eqn:acc}), we obtain the angular momentum of each halo.
In this equation, $\delta^{k}$ corresponds to the above averaged
density contrast $\delta_{avg}$.

\begin{figure}
\epsfxsize=8cm
\epsfbox{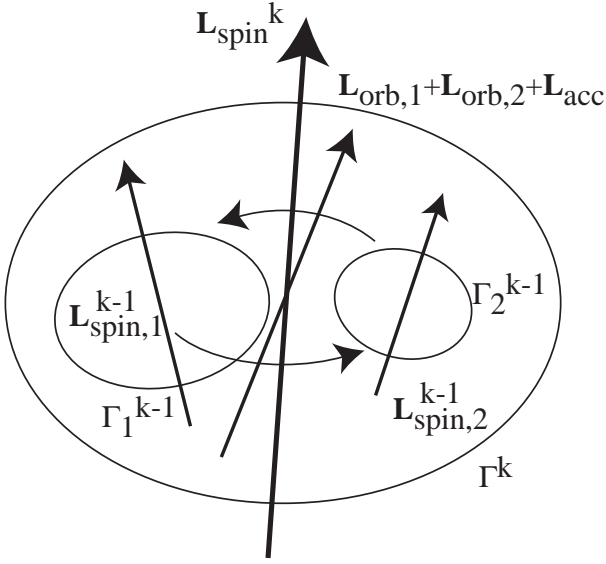}
\caption{Schematic description of dark halo formation including two
pre-existing haloes.  The halo labelled by the superscript $k$, which
is surrounded by the ellipsoid $\Gamma^{k}$, is formed through the
merger of the two progenitors labelled by $k-1$, which are surrounded
by the ellipsoids $\Gamma_{1}^{k-1}$ and $\Gamma_{2}^{k-1}$, and the
accretion of matter within $\Gamma^{k}$.  These two progenitors have
their spins $\bfL_{spin,1}^{k-1}$ and $\bfL_{spin,2}^{k-1}$, and their
orbital angular momenta around the centre-of-mass of the new common
$k$-th generation halo $\bfL_{orb,1}$ and $\bfL_{orb,2}$.  These
orbital angular momenta and the accreting matter's angular momentum
$\bfL_{acc}$ grow until the maximum expansion time of the new common
halo.  Thus the angular momentum of the new common halo is obtained by
$\bfL_{spin}^{k}=\bfL_{spin}+\bfL_{orb}+\bfL_{acc}$.}\label{fig:image}
\end{figure}

In Fig.\ref{fig:image}, we show the schematic description of dark halo
formation including two progenitors.  In this case, two $(k-1)$-th
generation progenitors merge together into a new $k$-th generation
larger halo with accreting matter.  So the spin component of the new
$k$-th generation halo is obtained as
$\bfL_{spin}=\bfL_{spin,1}^{k-1}+\bfL_{spin,2}^{k-1}$, and the total
angular momentum, namely, the spin of the new $k$-th generation halo
is $\bfL_{spin}^{k}=\bfL_{spin}+\bfL_{orb}+\bfL_{acc}$.

%%%%%%%%%%%%%%%%%%%%%%%% RESULT %%%%%%%%%%%%%%%%%%%%%%%%%%%
\section{RESULTS}
In this section we show final distributions of the spin angular
momentum $|\bfL|$ and the mass-weighted angular momentum
$|\bfL|/M^{5/3}$, which are similar to those used by Heavens \&
Peacock (1988) and Catelan \& Theuns (1996a).  We assume that linear
density fluctuations obey a random Gaussian distribution with a
power-law power spectrum $P(k)\propto k^{n}$, where $n=0$ and $-2$.
We consider only the Einstein-de Sitter background universe
($\Omega=1, \Lambda=0$).  The calculating box size is
$L_{\mbox{box}}=128$.  The normalization of the power spectrum is such
that the variance of the density contrasts equals unity at a block
with eight cells.  The overlapping parameters are set as $x=y=1/2$,
and $x=1/8$ and $y=1/2$.

The number of haloes finally obtained is about $1.6\times 10^{4}$ for
$n=0$ and about $9\times 10^{3}$ for $n=-2$ in one calculating box.
The following results are averaged over five realizations, therefore
our investigations are statistically robust.

In all figures, the displayed values of the mass, the angular momentum
and the time are normalised.  The time $t$ is normalized by the
present time $t_{0}$.  Physical values for the mass and the angular
momentum at the present epoch may be obtained through
$M_{phys}=m_{*}M$ and $\bfL_{phys}=m_{*}l_{*}\bfL$, where
$m_{*}=1.4\times 10^{14}M_{\odot}$ and $l_{*}=3\times 10^{6}
\mbox{km~s$^{-1}$~kpc}$ for the Hubble parameter $h=0.5$, which is
normalised by 100~km s$^{-1}$Mpc$^{-1}$ ($m_{*}$ corresponds to the
mass of one cell and $l_{*}$ is determined by the resolution of the
calculation).  Note that these physical values are scaled as
$m_{*}(z)=m_{*}(1+z)^{-6/(3+n)}$ because of the scale-free power
spectrum.  For example, $m_{*}\simeq 3\times 10^{9}M_{\odot}$ at $z=5$
in the case of $n=-2$.

In the following, we consider three cases in which to estimate the
angular momentum.  One is the merger case, in which ${\bf L}$ is given
by eq.(\ref{eqn:lnew}).  Second is the non-orbital case, in which we
remove the component of orbital angular momentum, ${\bfL}_{orb}$.
Third is the homogeneous case.  In the homogeneous case, we estimate a
region occupied by each halo, then we average the density contrasts in
the region; after that, we estimate the maximum expansion time via the
averaged density contrast by using eq.(\ref{eqn:max}).  Thus each cell
which belongs to a halo has the same maximum expansion time, that is,
the halo is treated as a homogeneous density object in the homogeneous
case.  By using this maximum expansion time, we obtain the angular
momentum of the final halo through the velocity field, as in
eq.(\ref{eqn:acc}).  This procedure is almost the same as that in the
previous work (Heavens \& Peacock 1988; Catelan \& Theuns 1996a, b).

In all cases, we consider only haloes with mass larger than nine
cells, because of the numerical resolution.  Also, we consider only
final haloes with an averaged density contrast larger than 1.69 in the
homogeneous case, because each collapsing block must satisfy the
criterion of the spherical collapse.

\begin{figure}
\epsfxsize=8cm
\epsfbox{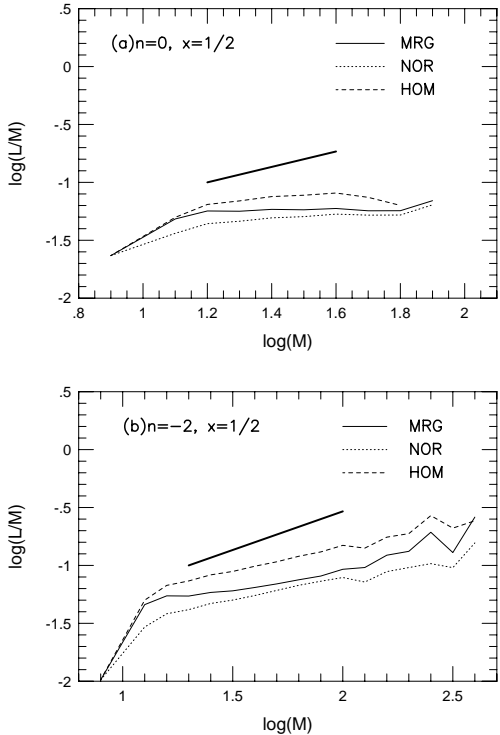}
\caption{Mass dependence of mean angular momentum in the case of
$x=1/2$.  upper panel: n=0, lower panel: n=-2.  The dashed lines show
the homogeneous case (HOM).  The solid lines show the merger case
(MRG).  The dotted lines show those without orbital angular momentum
(the non-orbital case; NOR).  The thick solid lines show the lines
proportional to $L\propto M^{5/3}$.}
\label{fig:spec1}
\end{figure}

\begin{figure}
\epsfxsize=8cm
\epsfbox{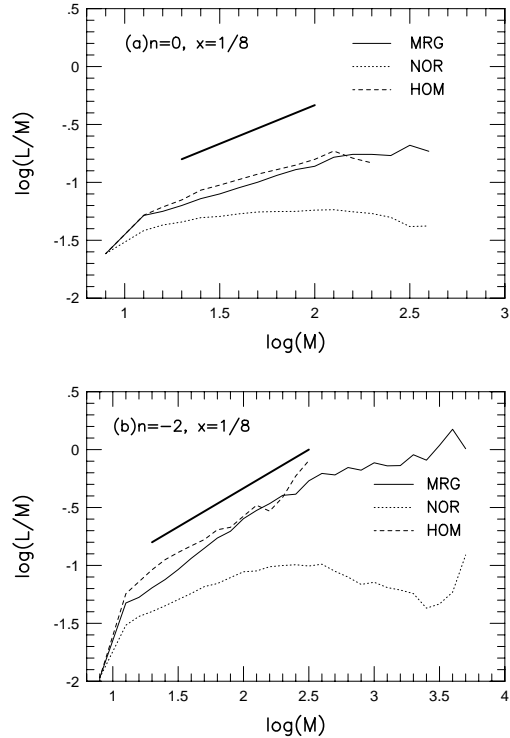}
\caption{The same as Fig.\protect\ref{fig:spec1} but for $x=1/8$.
upper panel: n=0, lower panel: n=-2.}
\label{fig:spec2}
\end{figure}

\subsection{Mass dependence of angular momentum}
First, we show the mean angular momentum against the mass of the
haloes in the case of $x=1/2$ (Fig.\ref{fig:spec1}) and $x=1/8$
(Fig.\ref{fig:spec2}).  The horizontal axis denotes the mass, that is,
the number of cells of each halo.  The vertical axis denotes angular
momentum averaged over each mass bin.  The solid lines correspond to
the merger case.  The dotted lines and dashed lines correspond to the
non-orbital case and the homogeneous case, respectively.  The short
thick solid lines denote $L\propto M^{5/3}$.  The minimum mass in the
figures corresponds to a halo with eight cells ($\log{8}\sim 0.9$),
and the values of the mean angular momentum in the minimum mass are
unphysically small because of the numerical resolution.  In the
homogeneous case, there are no data for larger-mass because the
averaged density contrasts in large haloes are smaller than 1.69.

The deviation from the thick line $L\propto M^{5/3}$ probably results
from the boundary effect of objects, caused by the artificial {\it
blocks} and {\it cells}.  We will show this boundary effect in a
future work.

In the case of $x=1/2$ (Fig.\ref{fig:spec1}), the differences among
the lines are rather small, but we find tendencies: the angular
momentum in the merger case is smaller than in the homogeneous case,
and the angular momentum decreases slightly when the orbital angular
momentum component is removed.  We find that orbital angular momentum
does not play an important role in the case of $x=1/2$.  In the case
of $x=1/2$, haloes grow almost solely through accretion events; large
blocks with $8^{3}$ or more cells should collapse in order to merge
two or more pre-existing haloes.  However, the probability that such
large blocks would collapse, including two or more haloes and
satisfying the criterion (c), is small.  Thus merger events between
two or more pre-existing haloes occur rarely.  This is the reason why
orbital angular momentum does not play an important role in the case
of $x=1/2$.

On the other hand, when $x=1/8$ we can see that the difference between
the merger case and the non-orbital case is large, about one order of
magnitude at $\log M\sim 3.5$, while the difference between the merger
case and the homogeneous case is not pronounced (Fig.\ref{fig:spec2}).
Under these conditions, the merger events occur easily.  Since the
orbital angular momentum can grow until a relatively later time, the
role of the orbital angular momentum becomes important.

In both cases, we find that the mass dependence of the homogeneous
case is almost the same as that of the merger case, while the
contribution of the orbital angular momentum to the total angular
momentum depends on the overlapping parameter $x$.  These properties
are almost independent of the spectral index $n$.  It seems to be
strange that the merger and the homogeneous cases lead to the same
results independently of the value of $x$.  We will investigate this
problem by considering the angular momentum distribution in detail in
the next subsection.

\subsection{Angular momentum distributions}

\begin{figure}
\epsfxsize=8cm
\epsfbox{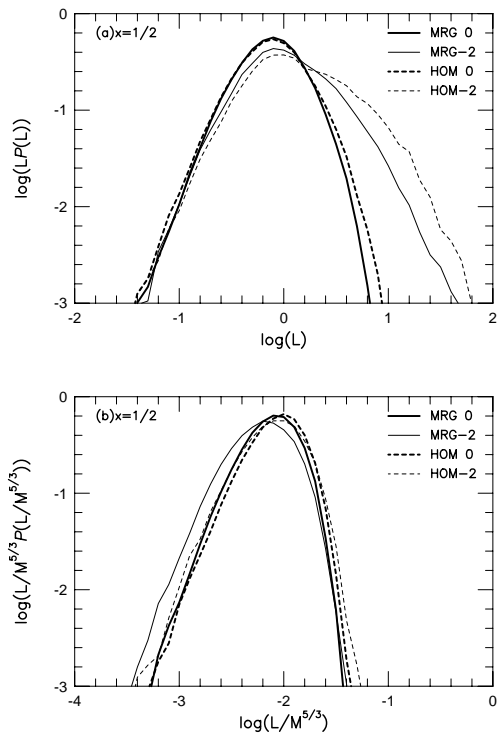}
\caption{(a) Angular momentum distributions in the case of $x=1/2$.
The thick lines denote the case of n=0, and the thin lines n=-2.  The
solid lines denote the merger case, and the dashed lines homogeneous
case.  (b) Mass-weighted angular momentum ($L/M^{5/3}$) distributions
in the case of $x=1/2$.  Types of the lines are the same as (a).}
\label{fig:dist1}
\end{figure}

\begin{figure}
\epsfxsize=8cm
\epsfbox{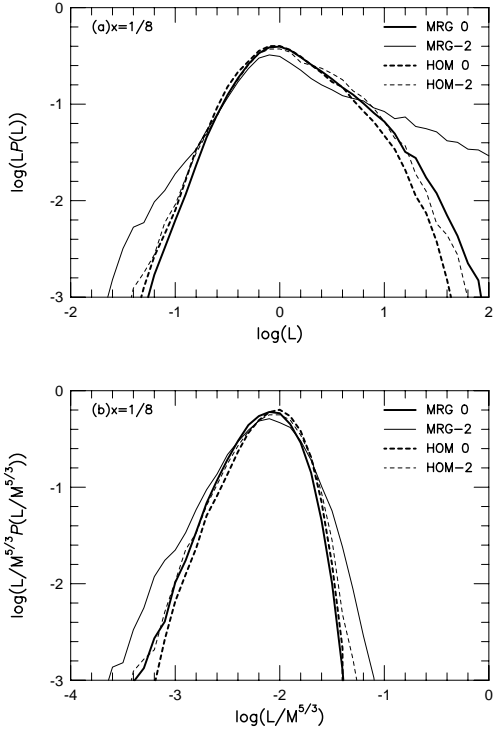}
\caption{The same as Fig.\protect\ref{fig:dist1}, but for
$x=1/8$.}\label{fig:dist2}
\end{figure}

To investigate the angular momentum distribution in detail, let us
consider the distributions of the angular momentum $L$ in
Fig.\ref{fig:dist1}a ($x=1/2$) and in Fig.\ref{fig:dist2}a ($x=1/8$),
and the mass-weighted angular momentum $L/M^{5/3}$ in
Fig.\ref{fig:dist1}b ($x=1/2$) and in Fig.\ref{fig:dist2}b ($x=1/8$).
Each figure includes the merger case and the homogeneous case.  The
solid lines show the distributions for the merger case, and the dashed
lines the homogeneous case.  The thick lines correspond to the
spectral index $n=0$, and the thin lines mark $n=-2$.

In the case of $x=1/2$, we can see that the distributions for the
homogeneous case are almost the same as those of the merger case.  The
degree of the difference between them is the same order of magnitude
as the difference between the mean values shown in
Fig.\ref{fig:spec1}.  This result suggests that the final distribution
of the angular momentum does not change very much when considering the
merger effect.

For $x=1/8$, it seems that the distributions of $L$ in the merger case
become wider especially in the case of $n=-2$ (Fig.\ref{fig:dist2}a).
It should be noted that haloes with averaged density contrast smaller
than 1.69 are removed in the homogeneous case, so the difference
between the homogeneous and the merger cases is apparently large at
higher angular momentum.  In Fig.\ref{fig:dist2}b, we do not need to
take care of this removal effect in the homogeneous case, because the
mean angular momentum is nearly proportional to $M^{5/3}$ (see the
thick solid line in Fig.\ref{fig:spec2}).  The distribution for the
homogeneous case is also almost the same as that for the merger case
when $x=1/8$.  We find that the merger effect does not affect the
final angular momentum distribution, independently of $x$.

Still, the difference between the merger case and the homogeneous case
is very small.  This fact is considered below in more detail.

\subsection{Distributions of maximum expansion time}
In this subsection, first we investigate the distribution of a product
of mass and the maximum expansion time of each halo, because the
absolute value of angular momentum depends strongly on the maximum
expansion time.  This quantity is represented as
\begin{eqnarray} 
M\langle
t_{m}\rangle&\equiv&\sum_{i}M_{cell}t_{m,i},
\end{eqnarray}
where $t_{m,i}$ is the first maximum expansion time of the cell
labelled by $i$, which is defined below in each model.  The sum is
carried out in the region of each halo.  So the angle brackets mean
the average in the region of each halo.  We distinguish the merger
case from the homogeneous case by superscripts `MRG' and `HOM', e.g.,
$\langle t_{m}^{MRG}\rangle$ and $\langle t_{m}^{HOM}\rangle$, because
values of $t_{m,i}$ are different in the case of the merger and the
homogeneous cases.  In the homogeneous case $t_{m,i}^{HOM}$ of a cell
is estimated from the density contrast averaged over the region of
each halo, therefore cells which belong to the same final halo have
the same maximum expansion time.  However, in the merger case
$t_{m,i}^{MRG}$ is estimated from the density contrast of a block to
which the cell belongs.

\begin{figure}
\epsfxsize=8cm 
\epsfbox{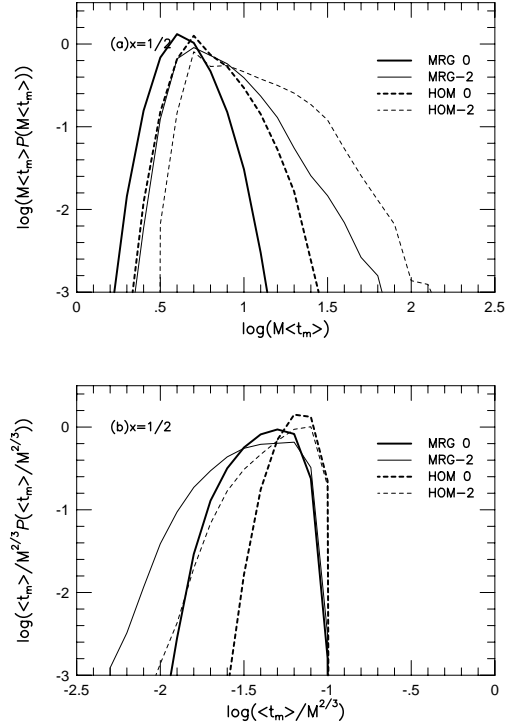}
\caption{Distributions of product of mass $M$ and maximum expansion
time $\langle t_{m}\rangle$ in the case of $x=1/2$.  (a) distributions
of $M\langle t_{m}\rangle$. (b) $\langle t_{m}\rangle/M^{2/3}$.  The
solid lines show the distributions for the merger case.  The dashed
lines show those for the homogeneous case.}\label{fig:time1}
\end{figure}

\begin{figure}
\epsfxsize=8cm
\epsfbox{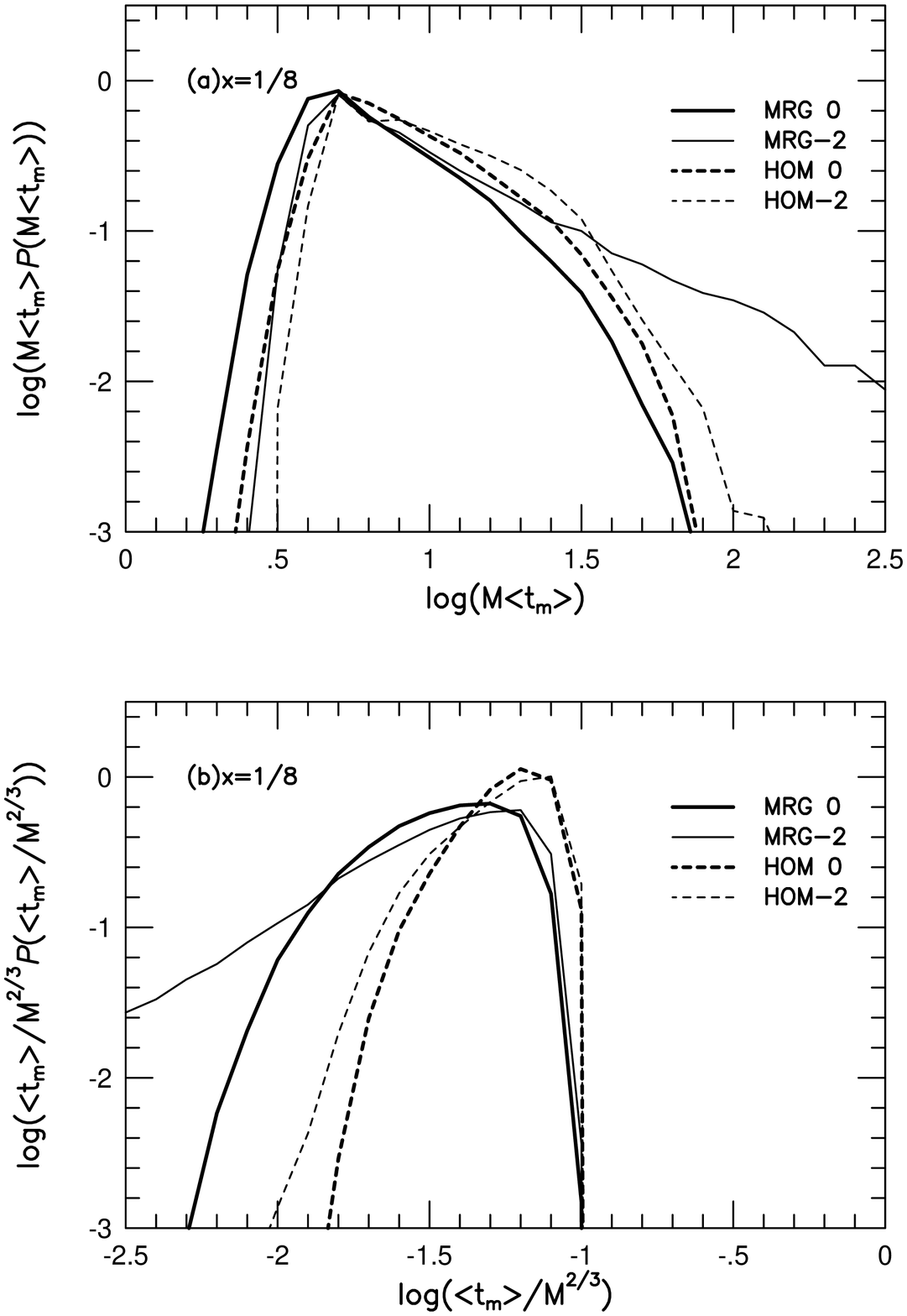}
\caption{The same as Fig.\protect\ref{fig:time1} but for
$x=1/8$.}\label{fig:time2}
\end{figure}

We show the distributions of $M\langle t_{m}\rangle$ in
Fig.\ref{fig:time1}a ($x=1/2$) and in Fig.\ref{fig:time2}a ($x=1/8$),
and $\langle t_{m}\rangle/M^{2/3}$ in Fig.\ref{fig:time1}b ($x=1/2$)
and in Fig.\ref{fig:time1}b ($x=1/8$), in the merger and the
homogeneous cases, respectively, because the dependences on the
maximum expansion time $L$ and $L/M^{5/3}$ in Figs.\ref{fig:dist1} and
\ref{fig:dist2} are the same as $M\langle t_{m}\rangle$ and $\langle
t_{m}\rangle/M^{2/3}$.  The solid lines show the distributions for the
merger case, and the dashed lines represent the homogeneous case.  The
thick lines and the thin lines mark the spectral index $n=0$ and
$n=-2$, respectively.

In the case of $x=1/2$, the differences between the merger case and
the homogeneous case are about 0.3 order of magnitude at
$\log(M\langle t_{m}\rangle)\sim 1 - 2$ (Fig.\ref{fig:time1}a) and at
the range of $\log(M\langle t_{m}\rangle)\sim -2.5 - -1.5)$
(Fig.\ref{fig:time1}b).  These differences are large as compared with
the distribution of the angular momentum shown in Fig.\ref{fig:dist1}.
This is caused by the numerous cells which halt the growth of the
angular momentum earlier in the merger case than in the homogeneous
case, especially for large haloes.  The result in the case of $x=1/8$
is the same as for $x=1/2$.  Therefore, we must explain the reason why
the distribution of the angular momentum is not changed much by the
merger effect, as compared to the distribution of the maximum
expansion time.

We should consider the effect of the direction of the angular
momentum.  If the growth of the angular momenta of the proto-objects
with a direction close to the direction of the final angular momentum
halts later than the growth of the angular momenta of the other
proto-objects with an opposite direction, in a region which becomes a
present-time halo, the total angular momentum of the halo can become
larger in spite of the averaged maximum expansion time $\langle
t_{m}^{MRG}\rangle$ being small.  Thus we investigate the maximum
expansion time, weighted by the direction (which will be defined
below).

Defining the time-independent vector part of the angular momentum of a
cell labelled by $i$ as
\begin{eqnarray} 
\bfL^{vec}_{i}=M_{cell}(\bfq_{i}-\bfq_{CM})\times(\bfv_{i}-\bfv_{CM}),
\end{eqnarray}
the absolute values of the angular momentum for the merger case and
for the homogeneous case are as follows
\begin{eqnarray}
|\bfL^{MRG}|&=&\left|\sum_{i}\bfL^{vec}_{i}t_{m,i}^{MRG}\right|,
\label{eqn:angmrg}\\ 
|\bfL^{HOM}|&=&\left|\sum_{i}\bfL^{vec}_{i}\right|
\frac{\sum_{i}M_{cell}t_{m,i}^{HOM}}{M}
=\left|\sum_{i}\bfL^{vec}_{i}\right|\langle t_{m}^{HOM}\rangle,
\label{eqn:anghom}
\end{eqnarray}
respectively.  Considering the merger case, defining $\theta_{f}$ as the
angle of the direction of the final angular momentum, we can transform
eqs. (\ref{eqn:angmrg}) and (\ref{eqn:anghom}) into 
\begin{eqnarray}
|\bfL^{MRG}|=\sum_{i}|\bfL^{vec}_{i}|\cos(\theta_{i}-\theta_{f})t_{m,i}^{MRG},
\\
|\bfL^{HOM}|\simeq\sum_{i}|\bfL^{vec}_{i}|\cos(\theta_{i}-\theta_{f})\langle 
t_{m}^{HOM}\rangle.
\end{eqnarray}
So we investigate the direction-weighted maximum expansion time
instead of $\langle t_{m}\rangle$,
\begin{eqnarray}
\langle
t_{m}\rangle_{\cos\theta}\equiv\frac
{\sum_{i}\cos(\theta_{i}-\theta_{f})t_{m,i}^{MRG}}
{\sum_{i}\cos(\theta_{i}-\theta_{f})}.
\end{eqnarray}
Then
\begin{eqnarray}
|\bfL^{MRG}|\simeq\sum_{i}|\bfL_{i}^{vec}|\cos(\theta_{i}-\theta_{f})\langle 
t_{m}\rangle_{\cos\theta}.
\end{eqnarray}
Here we have used the assumption that the absolute value of
$\bfL_{i}^{vec}$ is statistically independent of the direction of
$\bfL_{i}^{vec}$.  Thus if $\langle t_{m}^{HOM}\rangle\sim\langle
t_{m}\rangle_{\cos\theta}$ it is expected that $|\bfL^{HOM}|\simeq
|\bfL^{MRG}|$.

We show the relation of the direction-weighted maximum expansion time
$\langle t_{m}\rangle_{\cos\theta}$, the non-weighted maximum
expansion time $\langle t_{m}^{MRG}\rangle$, and the averaged maximum
expansion time $\langle t_{m}^{HOM}\rangle$ in Fig.\ref{fig:plot1}
($x=1/2$) and in Fig.\ref{fig:plot2} ($x=1/8$).  Each dot corresponds
to a halo.  The left panels show results for the spectral index $n=0$
and the right panels for $n=-2$.  We show the comparisons of $\langle
t_{m}^{HOM}\rangle$ with $\langle t_{m}\rangle_{\cos\theta}$ in the
upper panels, and with $\langle t_{m}^{MRG}\rangle$ in the lower
panels.  The solid straight lines have a unity slope.

The difference of the extents of the distributions of dots in the
horizontal axis between the case of $n=0$ and $n=-2$ is a result of
the difference of the power spectrum of density fluctuation field.
The mass dependence of the maximum expansion time is
\begin{eqnarray}
t_{m}\propto a_{M}^{3/2}=\left(\frac{1.05}{\delta}\right)^{3/2}\sim
M^{\frac{3+n}{4}},
\end{eqnarray}
thus the distribution of dots in the case of $n=0$ spread out wider
than those in $n=-2$.

In the lower panels, there are most of dots below the solid lines, as
we have already found in Figs.\ref{fig:time1} and \ref{fig:time2}.
However, in the upper panels, the dots are distributed in a wider
range in the vertical axis than those in the lower panels.  This
result shows that in some merging histories, the growth of angular
momentum with the opposite direction to the final one happens to halt
earlier than that with the same direction of the final one, and so
some haloes satisfy the condition $\langle
t_{m}\rangle_{\cos\theta}\geq\langle t_{m}^{MRG}\rangle$.  Then
$\langle t_{m}\rangle_{\cos\theta}\sim\langle t_{m}^{HOM}\rangle$
while the dispersion of $\langle t_{m}\rangle_{\cos\theta}$ is large.

This is the main reason why the distribution of the angular momentum
in the merger case is almost the same as that in the homogeneous case.
Although $\langle t_{m}^{MRG}\rangle$ is smaller than $\langle
t_{m}^{HOM}\rangle$, proto-objects having an angular momentum with the
same direction as the final one have a larger maximum expansion time
$t_{m,i}^{MRG}$, and those with the opposite direction have smaller
maximum expansion time.  So the final angular momentum becomes almost
the same as that for the homogeneous case.  Therefore both
distributions are very much alike.

\begin{figure}
\epsfxsize=8cm 
%\epsfbox{fig9.eps}
\caption{Distributions of maximum expansion time $\langle
t_{m}\rangle$ in the case of $x=1/2$.  The solid lines denote those
with a unity slope.  (a) Distributions of $\langle t_{m}^{HOM}\rangle$
(the homogeneous case) v.s. $\langle t_{m}\rangle_{\cos\theta}$ in the
case of $n=0$.  (b) Distributions of $\langle t_{m}^{HOM}\rangle$ (the
homogeneous case) v.s. $\langle t_{m}^{MRG}\rangle$ (the merger case)
in the case of $n=0$.  (c) The same as (a), but for $n=-2$. (d) The
same as (b), but for $n=-2$. }\label{fig:plot1}
\end{figure}

\begin{figure}
\epsfxsize=8cm
%\epsfbox{fig10.eps}
\caption{The same as Fig.\protect\ref{fig:plot1} but for
$x=1/8$.}\label{fig:plot2}
\end{figure}

\subsection{Orbital angular momentum}

\begin{figure}
\epsfxsize=8cm
\epsfbox{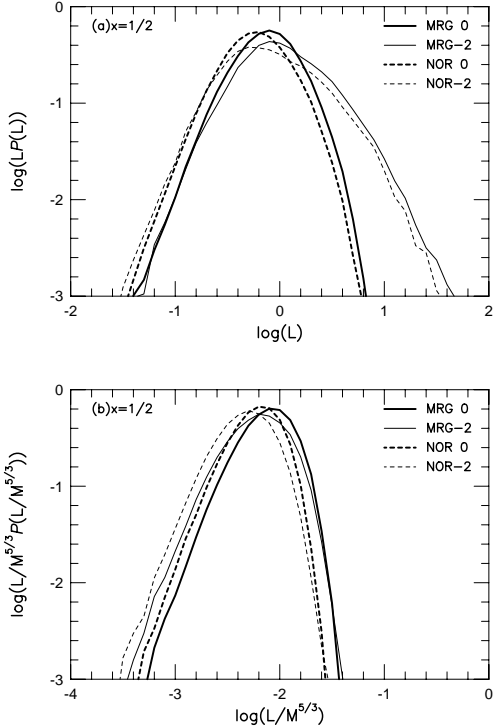}
\caption{(a) Angular momentum distributions in the case of $x=1/2$.
The thick lines denote the case of $n=0$, and the thin lines $n=-2$.
The solid lines are the same as Fig.\protect\ref{fig:dist1} (the
merger case), and the dashed lines denote the non-orbital case
(angular momentum without the orbital angular momentum).  (b)
Mass-weighted angular momentum $(L/M^{5/3})$ distributions in the case
of $x=1/2$.  Types of the lines are the same as (a).}\label{fig:orb1}
\end{figure}

In order to investigate the contribution of the orbital angular
momentum to the total angular momentum, in Fig.\ref{fig:orb1} we show
the case in which we remove the component of the orbital angular
momentum (the non-orbital case) in the case of $x=1/2$.  The solid
lines show the total angular momentum
($\bfL_{spin}+\bfL_{acc}+\bfL_{orb}$) in Fig.\ref{fig:orb1}a and those
divided by $M^{5/3}$ in Fig.\ref{fig:orb1}b for the merger case, and
the dashed lines show the total angular momentum without the orbital
angular momentum component, that is, $\bfL_{spin}+\bfL_{acc}$ in
Fig.\ref{fig:orb1}a and those divided by $M^{5/3}$ in
Fig.\ref{fig:orb1}b.  The thick lines denote the spectral index $n=0$,
and the thin lines $n=-2$.  In the case of $x=1/2$, reflecting the
small differences in Fig.\ref{fig:spec1}, the differences between the
merger case and the non-orbital case are very small.  As already
mentioned in Section 5.1, in the case of $x=1/2$, haloes grow almost
solely through the accretion events, and the merger events between two
or more pre-existing haloes occur rarely.  Thus the contribution of
the orbital angular momentum is small.

\begin{figure}
\epsfxsize=8cm
\epsfbox{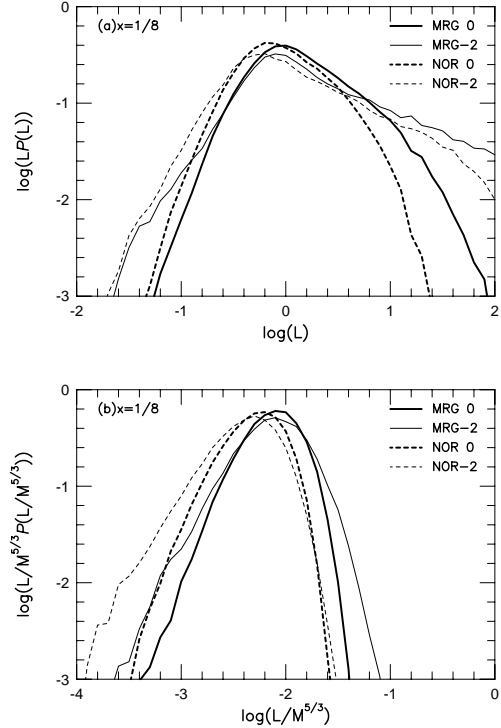}
\caption{The same as Fig.\protect\ref{fig:orb1} but for
$x=1/8$.}\label{fig:orb2}
\end{figure}

In Fig.\ref{fig:orb2}, we show the same results as in
Fig.\ref{fig:orb1} but for the case of $x=1/8$.  In the upper panel,
in contrast to Fig.\ref{fig:orb1}a, the differences at higher angular
momentum are larger than those at lower angular momentum.  In the case
of $x=1/8$, two or more pre-existing haloes can merge with each other
many times.  We find that the number of occurrences of the merger
event is nearly proportional to $0.01M^{1.25}$ in the case of $n=0$
and $0.02M$ in the case of $n=-2$ for larger haloes than $\sim 100$
cells.  Then haloes with a larger mass generally experience many
merging events.  Moreover, the haloes with larger mass have larger
angular momentum.  Hence the contribution of the orbital angular
momentum of a halo is greater for the halo with a larger total angular
momentum.

\subsection{Spin parameter $\lambda$}

In this subsection, we show the distributions of the dimensionless spin
parameter $\lambda$, which is defined as
\begin{equation}
\lambda\equiv\frac{L|E|^{1/2}}{GM^{5/2}},
\end{equation}
where $E$ is the total energy of a halo with mass $M$ and angular
momentum $L$.  We estimate $E$ by summing kinetic energies and
gravitational energies of cells within each halo.

This spin parameter corresponds to the ratio of the rotational kinetic
energy to the gravitational energy, so an object with $\lambda\sim 1$
is fully rotationally supported.  It has been reported that typically
$\lambda\approx 0.08$ for the theoretical predictions and for the
N-body simulations, and $\lambda\sim 0.5$ for spirals and $\lambda\sim
0.05$ for ellipticals by observations.  We need to know dissipational
gas collapse process in order to compare the spin parameters obtained
theoretically with those obtained by observations, but it is beyond
the scope in this paper.

\begin{figure}
\epsfxsize=8cm
\epsfbox{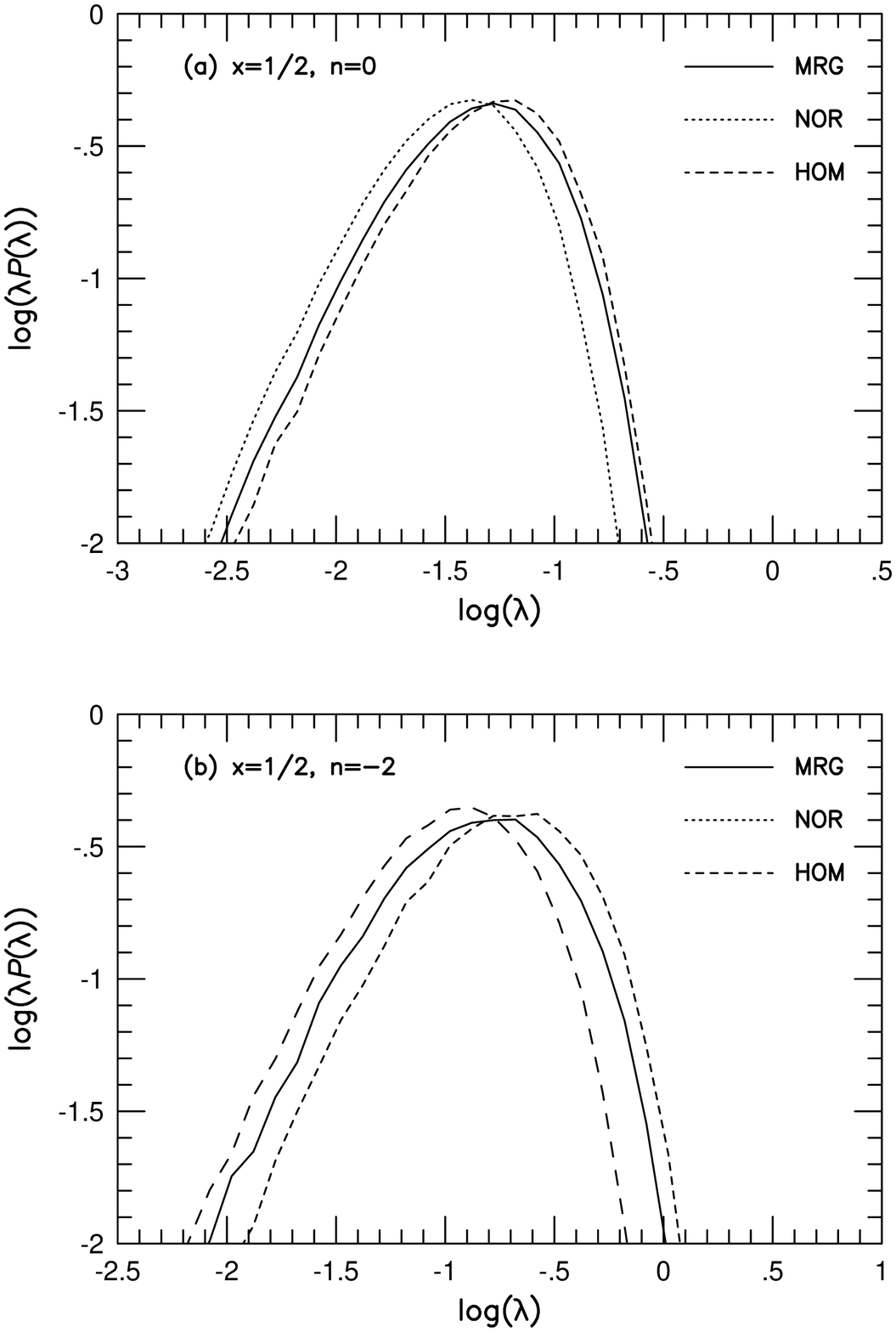}
\caption{$\lambda$ distributions in the case of $x=1/2$.  (a)
$n=0$. (b) $n=-2$.  The solid lines are the merger case, the dashed
dashed lines denote the homogeneous case, and the dotted lines show
the non-orbital case.}\label{fig:lam1}
\end{figure}

\begin{figure}
\epsfxsize=8cm
\epsfbox{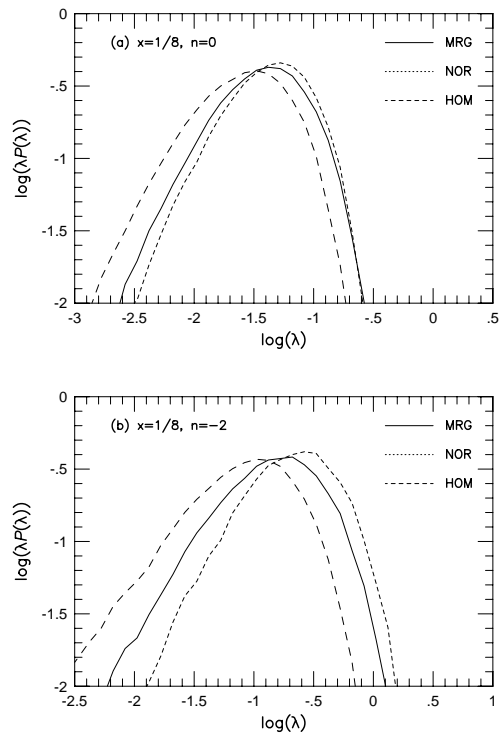}
\caption{The same as Fig.\protect\ref{fig:lam1} but for
$x=1/8$.}\label{fig:lam2}
\end{figure}

In Fig.\ref{fig:lam1}, we show the distributions of $\lambda$ in the
merger (solid lines), homogeneous (dashed lines), and non-orbital
cases (dotted lines) when $x=1/2$.  Similarly, in Fig.\ref{fig:lam2},
we show the same lines as Fig.\ref{fig:lam1} in $x=1/8$.  In both
cases, the values of $\lambda$ at the peaks of distributions in the
merger case are lower than those in the homogeneous case, and are
higher than those in the non-orbital case.  These tendencies are the
same as the results in the previous subsections.  The rms values of
$\lambda$ are $\sqrt{\langle\lambda^{2}\rangle}\simeq 0.07$ in the
case of $n=0$ and $0.25$ in the case of $n=-2$ in the merger case.
These are almost consistent with the previous theoretical predictions.

%%%%%%%%%%%%%%%%%%%%%%% section ``conclusions & discussion'' %%%%%%%%%%  
\section{CONCLUSIONS AND DISCUSSION}  
In this paper we have analysed the acquisition and distribution of the
angular momentum of proto-objects with inhomogeneous density
distribution by using the merging cell model.  The assumptions in our
calculations are as follows: spherically symmetric collapse of each
block, the overlapping criteria (see Section 3), and the growth of
angular momentum proportional to the cosmic time $t$.  We also assume
that the halting time of the acquisition of angular momentum is
determined by the maximum expansion time.

In almost all of the previous work, it has been suggested that
proto-objects are homogeneous density ellipsoids, and that the maximum
expansion time may be estimated from the smoothed density contrast.
However, if we consider the hierarchical clustering scenarios on which
most of the previous analyses of the angular momentum are based, we
should consider the merger effect.  In the hierarchical clustering
scenario, it is natural that each proto-object has a different halting
time in the acquisition of the angular momentum.  Hence it is very
important to consider the merging histories of dark haloes.  However,
we find no significant differences between the distributions of
angular momentum for the merger and the homogeneous cases in our
analyses by using the merging cell model.  We find the reason for this
by dividing the angular momentum into two components, that is, the
time-independent vector and the maximum expansion time (see Section
5.3).  The distribution of the maximum expansion time, which is
directly influenced by the merging history, moves to the lower value.
However, the angular momentum components that have a direction which
is the same as that of the final angular momentum grow later than
those with other directions.  This is shown by comparing simply
averaged maximum expansion time in each halo $\langle
t_{m}^{MRG}\rangle$ with the direction-weighted maximum expansion time
$\langle t_{m}\rangle_{\cos\theta}$.  We find that many haloes satisfy
$\langle t_{m}\rangle_{\cos\theta}\geq\langle t_{m}^{MRG}\rangle$.
Thus, as these two effects cancel each other out, the difference
between the merger case and the homogeneous case becomes negligible.
Note that this conclusion is independent of the overlapping parameter
$x$.

In our calculations, we use the scale-free power spectra, $P(k)\propto
k^{n}$, where $n=0$ and $n=-2$.  When we normalize $P(k)$ at the
present epoch, the mass-scale in our investigation corresponds to the
clusters of galaxies.  However, we can scale the normalization to a
galactic scale at high redshift because of the scale-free spectra (see
Section 5).  When we consider the galaxy formation processes, it may
be important to analyse the process of the acquisition of the angular
momentum of dark haloes at high redshift corresponding to the galaxy
formation epoch, because the spins of dark haloes convert to those of
the baryonic component, namely, galaxies at high redshift.

We also show the contributions of the orbital angular momentum of
pre-exiting haloes when mergers occur.  This analysis is enabled by
using the semi-analytical models like the merging cell model.  The
difference between the merger case and the non-orbital case, which
does not include the orbital angular momentum component strongly
depends on the overlapping parameter $x$.  In the case of $x=1/2$, in
which there are not too many merger events between two or more
pre-existing haloes, the contribution of the orbital angular momentum
to the total angular momentum is small.  On the other hand, in the
case of $x=1/8$, the orbital angular momentum plays an important role.
The contribution of the orbital angular momentum is about one order of
magnitude larger than the contributions of the other components, which
are the spin angular momentum and the angular momentum of the
accreting matter.  Note that it is easy for blocks to collapse in this
case of $x=1/8$ and that there are many merger events.  We cannot
decide the value of the overlapping parameter $x$ without a detailed
study of the overlapping condition, so we can only mention our
conclusions qualitatively.  However, even under this limitation, we
find that the orbital angular momentum plays a significant role in the
total angular momentum if merger events occur many times.  When the
evolution of the angular momentum of the baryonic component is
considered, the process of the merging of galaxies as gaseous and
stellar components will involve a release of angular momentum to dark
matter and running away gas.  So the distribution of the angular
momentum of the gaseous component may follow the distribution without
the orbital angular momentum.  If elliptical galaxies are formed
through such a merger process, and if spiral galaxies are formed
without a loss of angular momentum of accreting gas, this result is
very suggestive, because the observational difference between the
angular momentum of ellipticals and spirals is also about one order of
magnitude (see {\it e.g.} Fall 1983 or Fig.9 in Catelan \& Theuns
1996a).  Such tendencies are presented in Fig.\ref{fig:spec2}.  We
need hydro-dynamical simulations in order to decide whether the
distribution of the angular momentum of galaxies is followed by the
distribution that includes orbital angular momentum or not, since our
model considers only dark matter component.  One such baryonic effects
has been discussed as the `angular momentum catastrophe' in Weil, Eke
\& Efstathiou (1998).

As mentioned above, we assume the spherically symmetric collapse of
each block in the merging cell model.  For this reason, we need the
overlapping parameter $x$.  So we believe that we need a new model
which includes nonspherical collapse in order to discuss the merger
effect quantitatively.  This study is in progress.

\section*{ACKNOWLEDGMENTS}    
We wish to thank Misao Sasaki and Satoru Ikeuchi for useful
suggestions, and Michihiro Yoshida for stimulating discussion.  We are
grateful to Sergei Levshakov for carefully reading part of the
manuscript of our paper.  We also thank the anonymous referee whose
comments led us to a substantial improvement of our paper.  This work
was supported in part by Research Fellowships of the Japan Society for
the Promotion of Science for Young Scientists (No.2265).  The
calculations were performed in part on VPP300/16R and VX/4R at the
Astronomical Data Analysis Center of the National Astronomical
Observatory, Japan.

\bsp
\end{document}